# Image transport through a disordered optical fiber mediated by transverse Anderson localization

Salman Karbasi[1], Ryan J. Frazier[1], Karl W. Koch[2], Thomas Hawkins[3], John Ballato[3], and Arash Mafi[1]*

[1] *Department of Electrical Engineering and Computer Science, University of Wisconsin-Milwaukee, Milwaukee, WI 53211, USA.*
[2] *Optical Physics and Networks Technology, Corning Incorporated, SP-AR-01-2, Sullivan Park, Corning, NY 14831, USA.*
[3] *Center for Optical Materials Science and Engineering Technologies (COMSET) and the Department of Materials Science and Engineering, Clemson University, Clemson, SC 29625, USA.*
*Corresponding author: mafi@uwm.edu*

**Transverse Anderson localization of light allows localized optical-beam-transport through a transversely-disordered and longitudinally-invariant medium. Its successful implementation in disordered optical fibers recently resulted in the propagation of localized beams of radii comparable with that of conventional optical fibers. We present what is, to the best of our knowledge, the first demonstration of optical image transport using transverse Anderson localization of light. The image transport quality obtained in the polymer disordered optical fiber is comparable with or better than some of the best commercially available multicore image fibers with less pixelation and higher contrast. It is argued that considerable improvement in image transport quality can be obtained in a disordered fiber made from a glass matrix with near wavelength-size randomly distributed air-holes with an air-hole fill-fraction of 50%. Our results open the way to device-level implementation of the transverse Anderson localization of light with potential applications in biological and medical imaging.**

Anderson localization is the absence of diffusive wave propagation in certain disordered media [1–4]. Transverse Anderson localization was first introduced by Abdullaev, et al., [5] and De Raedt, et al., [6]. In the treatment suggested by De Raedt, et al., an optical wave system is studied in which the refractive index profile is random in the transverse plane and is invariant in the longitudinal direction. An optical beam launched in the longitudinal direction can become localized in the transverse plane due to the strong random scattering from the transverse random index fluctuations and can propagate in the longitudinal direction with a finite beam radius, similar to a conventional optical fiber. The radius of the localized beam depends on the extent of the fluctuations in the value of the refractive index, the characteristic length over which the fluctuations occur, as well as the wavelength of the light [7–9].

Transverse Anderson localization was first observed experimentally in 2007 [10]. That experiment, carried out in a photo-refractive crystal, utilized refractive index variations of the order of $10^{-4}$. Such small variations in the refractive index of random sites result in a very large mean value and standard deviation of the localized beam radius [9]. For device applications that benefit from the waveguiding properties of transverse Anderson localization, the mean localized beam radius should be comparable with or smaller than that of conventional optical fibers and large sample-to-sample variations in the beam radius are not acceptable. Therefore, it is necessary to increase the refractive index difference of the random sites.

We recently demonstrated transverse Anderson localization of light in a polymer random optical fiber medium with refractive index fluctuations of order 0.1 [7]. The polymer Anderson localized fiber (p-ALOF) allowed for the simultaneous propagation of multiple beams in a single strand of disordered optical fiber [8] with potential applications in beam-multiplexed optical communications and optical imaging.

In this work, we present what is, to the best of our knowledge, the first demonstration of optical image transport using transverse Anderson localization of light, specifically in a disordered optical fiber. The possibility of using disordered optical fiber for some form of image transport was expected, given the earlier demonstration of spatial beam multiplexing in p-ALOF [8]. The novelty of the presented work is in demonstrating that the image transport quality can be of a comparable or higher quality than the commercially available multicore imaging optical fibers. It is remarkable that the high quality image transport is achieved because of, not in spite of, the high level of disorder and randomness in the imaging system.

Multicore optical fibers have been used extensively in high resolution optical imaging [11]; however, the transmitted images are inherently pixelated due to the discrete nature of the light-guiding array of cores, and the inter-core coupling can reduce the image contrast and result in blurring [12, 13]. Certain structural non-uniformities such as variations in the size of the cores were shown by Ref. [14] to improve the image transport quality. A weakly disordered fiber array was also studied in Ref. [15] and was shown to induce diffusive spreading or localization at a few sites across the fiber. High numerical aperture guiding cores were also suggested by Ref. [13] to reduce core-to-core coupling and blurring in imaging applications.

A highly disordered optical fiber with large refractive index fluctuations can transport high quality images, as it provides a high degree of structural non-uniformity as well as a sufficiently large local numerical aperture. More rigorously, the image transport quality is due to the transverse Anderson localization phenomenon that creates localized transport channels with finite radii (localized optical modes) through the disordered imaging waveguide [16, 17]. A higher amount of disorder and a larger level of fluctuation in the refractive index provides stronger beam localization, hence improving the image resolution. It is also responsible for the reduction in the value of the standard deviation in the localized beam radius as a consequence of the self-averaging behavior [7, 18, 19], ensuring uniform image transport quality across the fiber facet. The coherent transverse coupling and blurring is considerably reduced, because the transverse disorder results in strong spatial incoherence across the beam, akin to using incoherent light to eliminate speckles in an imaging system. Therefore, even a laser can be readily used for illumination in this image fiber to obtain a higher signal-to-noise ratio, without worrying about its undesirably high spatial coherence.

The fabrication of the p-ALOF used in these imaging experiments is described in detail in Refs. [7, 20]; briefly, it is composed of 40,000 strands of poly methyl methacrylate (PMMA) and 40,000 strands of poly styrene (PS) drawn to a square profile with a side width of 250-$\mu m$ and site sizes of about 0.9-$\mu m$. A magnified scanning electron microscope (SEM) image of a portion of the tip of the p-ALOF is shown in Fig. 1(a) where the PMMA sites (refractive index of 1.49) are darker compared with the lighter PS sites (refractive index of 1.59). Also shown in Fig. 1(b), for comparison, is an SEM image of a portion of the tip of a glass disordered fiber earlier reported in Ref. [21], where the darker sites are the airholes. In the following, we demonstrate high-quality optical image transport through the p-ALOF related to Fig. 1(a) mediated by transverse Anderson localization of light. We also argue that a higher disorder density is required for quality image transport through the glass disordered fiber of Fig. 1(b) and a higher airhole fill-fraction of nearly 50% should result in a very high-quality image fiber.

**Results**

**Image transport in the disordered fiber.** In order to investigate the image transport capability of the p-ALOF, optical images of the "numbers" from group 3 and group 4 on an 1951 U.S. Air Force resolution test chart (1951-AFTT), Fig. 2, are launched using a 405 nm laser diode into the p-AOLF (see Methods). Different numbers in each group are the same size. The transported images for the groups 3 and 4 are shown in Fig. 3 and Fig. 4, respectively, after 5 cm of propagation in the p-ALOF. Theoretically, the minimum resolution of the images is determined by the width of the point spread function of the p-ALOF imaging setup, which is comparable with the localization length. The transverse localization length (localized beam radius) of the p-ALOF at 405 nm wavelength was calculated to be smaller than 10 $\mu m$ [9]. In practice, the imaging resolution in p-ALOF is limited by the quality of the cleave and polishing of the p-ALOF surface. The fiber surface quality both at the input and output is partially responsible for the distortions observed in the transported images in Figs. 3 and 4.

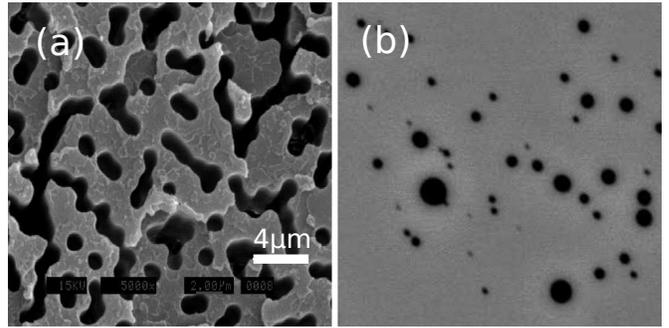

Fig. 1. **SEM images of the disordered fibers.** Magnified SEM image of a portion of the tip of the (a) p-ALOF and (b) glass disordered fiber. For p-ALOF in (a), the PMMA (PS) sites are darker (lighter) in color. For the glass disordered fiber, the darker sites are the airholes. The 4 $\mu m$ scale-bar applies to both images.

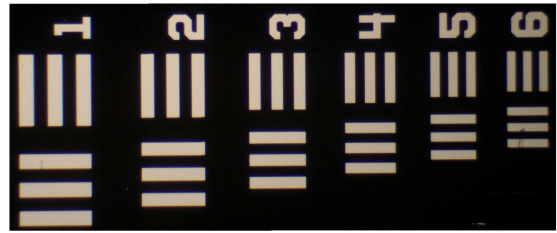

Fig. 2. **Elements of a group on 1951 U.S. Air Force test target (1951-AFTT).**

**Comparison with commercial multicore image fibers.** The imaging performance of the p-ALOF compares with some of the best commercially available multicore imaging optical fibers, as shown in Fig. 5. The transported images over 5 cm of the number "6" from group 5 of the 1951-AFTT test chart are compared between p-ALOF in Fig. 5(a), Fujikura FIGH-10-350S in Fig. 5(b), and Fujikura FIGH-10-500N in Fig. 5(c). The visual image quality of the transported image through the p-ALOF is clearly better than FIGH-10-350S and is comparable with FIGH-10-500N. We would like to emphasize that the feature sizes in Fig. 5 are of the order of 10-20 $\mu m$. The Rayleigh range for this level of resolution and the 405 nm laser wavelength is approximately 0.8-3 mm, which is substantially shorter than the 5 cm propagation length in these image fibers. Therefore, the imaging results are non-trivial and cannot be obtained using bulk propagation or conventional multimode opti-cal fibers. In order to quantify the perceived image quality of the transported images, we use the mean structural similarity index (MSSIM) throughout this manuscript. Details and justifications on the use of this metric are given in the Methods section and the Supplementary Information. The MSSIM values are 0.5877 for Fig. 5(a), 0.5501 for Fig. 5(b), and 0.5591 for Fig. 5(c).

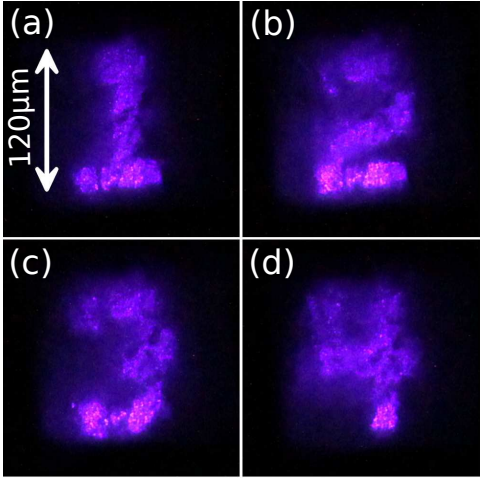

Fig. 3. **Transported images through the disordered fiber.** Transported images of different numbers through 5 cm of p-ALOF, (a)-(d), related to the group 3 on the 1951-AFTT (experimental measurements).

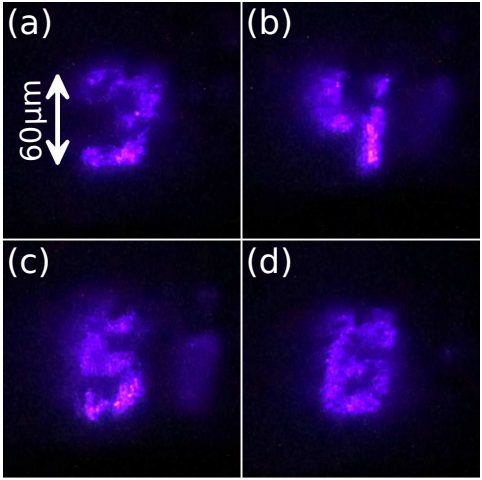

Fig. 4. **Transported images through the disordered fiber.** Transported images of the numbers through 5 cm of p-ALOF, (a)-(d), related to the group 4 on the 1951-AFTT (experimental measurements).

Using the optical and physical parameters of the p-ALOF, Fujikura FIGH-10-350S, and Fujikura FIGH-10-500N, we repeat the experiment of Fig. 5 using the numerical simulation and show the results in Fig. 6(a), Fig. 6(b), and Fig. 6(c). The numerical simulation is detailed in the Methods section. Similar to Fig. 5, the transported images are over 5 cm of the number "6" created using the GIMP image editor and nearly the same size as the group 5 of the 1951-AFTT test chart and are compared between p-ALOF in Fig. 6(a), Fujikura FIGH-10-350S in Fig. 6(b), and Fujikura FIGH-10-500N in Fig. 6(c).

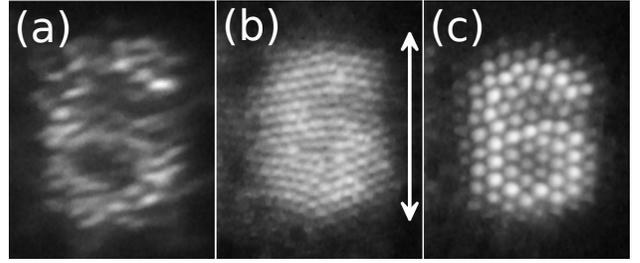

Fig. 5. **Transported images through the disordered fiber and the commercial image fibers.** Transported images related to group 5 of the 1951-AFTT test chart in (a) p-ALOF, (b) FIGH-10-350S image fiber and (c) FIGH-10-500N image fiber (experimental measurements). The scale bar in (b) is 30 $\mu$m-long and the same scale bar can be used for (a) and (c). Each fiber is approximately 5 cm long. The MSSIM image quality values for the images are: (a) 0.5877, (b) 0.5501, and (c) 0.5591.

Similar to the experimental results of Fig. 5, the numerical simulation shown in Fig. 6 indicates that the visual image quality of the transported image through the p-ALOF is better or comparable with the Fujikura image fibers. The experiment and numerics are in reasonable agreement, but there are differences, as well. Possible sources of difference can be traced back to uncertainties in relating the experiment to numerics. The MSSIM values are 0.637 for Fig. 6(a), 0.615 for Fig. 6(b), and 0.6257 for Fig. 6(c).

**Sources of systematic image degradation.** In the experiment, a good precision in the butt-coupling of the test target to the input fiber is needed to obtain high-quality output images. An important source of uncertainty is the surface quality of the fibers, determining the precision in coupling the input profile from the test target and coupling the output to the CCD camera. The Fujikura image fibers are cleaved and polished using commercial-grade equipment, while the p-ALOF is hand-cleaved and polished using the lapping paper from Thorlabs. These variations and uncertainties cannot be easily accounted for in the numerical simulation without extensive surface quality characterizations of p-ALOF and Fujikura image fibers and are likely not very illuminating, considering that the p-ALOF images can be improved if specialized equipment for cleaving and polishing polymer fibers are used to improve its surface quality. Another possible source of uncertainty is the degree of spatial coherence of the laser used to illuminate the test target. The spectral bandwidth of the laser also contributes the fuzziness of the experiment, while the single-frequency numerical simulation looks more pixelated.

We note that the scalar wave equation has been bench-marked extensively with the full vectorial beam propagation method and the results have shown to be in excellent agreement for the numerical apertures relevant in these simulations [9].

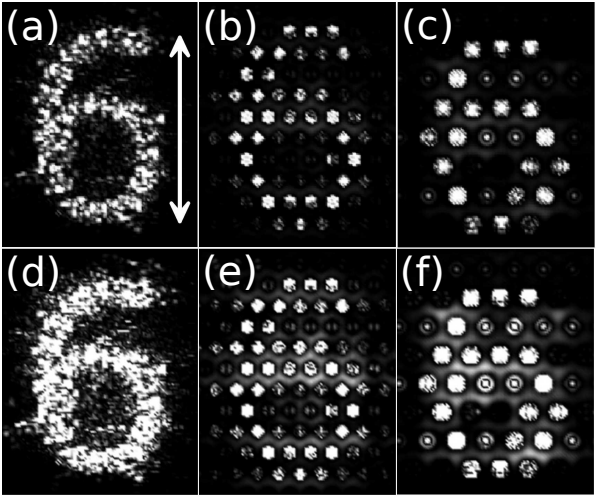

Fig. 6. **Simulation of image transport through the disordered fiber and the commercial image fibers.** Transported images related to group 5 of the 1951-AFTT test chart in (a) p-ALOF, (b) FIGH-10-350S image fiber and (c) FIGH-10-500N image fiber (numerical simulations). The scale bar in (a) is 30 $\mu$m-long and the same scale bar can be used for other subfigures. Each fiber is approximately 5 cm long. The top row (subfigures (a), (b), and (c)) can be compared with the experimental results shown in Fig. 5. Subfigures (d), (e), and (f) in the bottom row are the same as Subfigures (a), (b), and (c) in the top row, except the images are saturated by changing the color axis in gray-scale colormap in Matlab. The color axis of [0,1] in subfigures (a), (b), and (c) is changed to color axis of [0,0.3] in (d), (e), and (f). The MSSIM image quality values for the images are: (a) 0.637, (b) 0.615, and (c) 0.6257. The MSSIM values for (d), (e), and (f) are the same as (a), (b), and (c), respectively.

The quality of the imaging setup and the saturation of the CCD camera versus the saturation of the images generated using numerics can be another source of perceived difference between the experiment and numerics. The images of Fig. 6(a), Fig. 6(b), and Fig. 6(c) are plotted using the color axis of [0,1] in gray-scale colormap in Matlab. In Fig. 6(d), Fig. 6(e), and Fig. 6(f), the color axis is changed to [0,0.3] for the same simulation, resulting in a more fuzzy image. It should also be noted that the numerical simulations are for images at the tip of the fiber, while those from the experiment are after passing through the near-field imaging setup discussed in the Methods.

The simulated images of multicore image fibers in Figs. 6 look quite different from the experiment in Fig. 5, even when one takes into account the uncertainties mentioned above. This issue was pointed out in Ref. [14], where it was suggested that the discrepancy between theory and experiment is because the image fibers are not composed of identical cores. The published core size specification is likely the average value with potentially a large standard deviation. While it is possible to build a concrete model to investigate this issue in detail, it is beyond the scope and interests of this manuscript.

**Disorder-induced localization is responsible for enhanced image transport.** The discussion so far has mainly focused on the perceived quality of the images in Fig. 5 and Fig. 6. Because the intention of this manuscript is to argue that transverse Anderson localization induced by the disorder results in a higher quality image transport when compared with multicore image fibers, it is important to ensure that similar specifications are used for the disordered and multicore image fibers to ensure that the comparison is conducted fairly. The core and cladding refractive indices of the Fujikura image fibers are 1.5 and 1.446, respectively, for which the index difference of 0.054 is lower than that of p-ALOF (0.1). In order to isolate the effect of disorder and Anderson localization on the image transport quality and eliminate the differences due to the numerical apertures, an image fiber with the same numerical aperture as p-ALOF and the structural parameters of FIGH-10-500N is modeled using the numerical simulation. The input image is the word "UWM," where each letter is 44-$\mu$m high, the lines in the letters are 6-$\mu$m thick, and the letters are separated by 26-$\mu$m. The intensity profiles of the transported images at the wavelength of 405 nm after 5 cm of propagation in p-ALOF and the image fiber with a raised numerical aperture are compared in Fig. 7(a) and 7(b). The quality of the transported image in Fig. 7(a) is higher than that of the modified image fiber in Fig. 7(b). There is considerable pixelation in Fig. 7(b) and a low intensity halo of illuminated cores fills the area in between the main lines of the transported letters. It appears that even with the same level of index difference, p-ALOF provides a comparable or better quality image transport compared with the multicore image fiber. It should be noted that the images in Fig. 7 appear to be of higher quality compared with Figs. 5 and 6, simply because they are of different sizes and resolutions. The MSSIM values are 0.8923 for Fig. 7(a) and 0.6263 for Fig. 7(b).

As further concrete evidence that the disorder is responsible for the higher image transport quality, the quantitative MSSIM metric for image quality assessment is compared in Fig. 8 for a collection of multicore image fibers with and without disorder. The blue squares in Fig. 8 represent the value of MSSIM for image transport through 5 cm of disorder-free ($\Delta = 0$) periodic image fiber as a function of the periodicity $\Lambda$. The radius of each individual core is 1.45 $\mu m$, which is the same as the mean radius of the cores in the FIGH-10-500N imaging fiber. The refractive index difference between the cores and common cladding is 0.1. The image that is transported through the image fiber is the number "6" created using the GIMP image editor and is 65 $\mu m$ high.

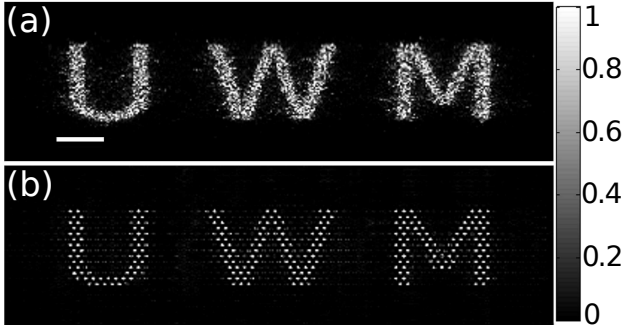

Fig. 7. **Simulation of image transport through the disordered fiber and an image fiber.** The intensity profile of the "UWM" image after 5 cm of propagation and at the wavelength of 405 nm in a (a) p-ALOF, (b) modified image fiber with the refractive index difference of 0.1 between the cores and the clad, using numerical simulations. The scale bar in (a) is 20 $\mu m$ long and the same scale bar can be used for (b). The MSSIM image quality value for the images are: (a) 0.8923 and (b) 0.6263.

In the case of disorder-free periodic image fiber, MSSIM is low for small $\Lambda$ as expected because of the enhanced core-to-core coupling. It is also low for large $\Lambda$ due to the pixelation effect. In general, the optimum $\Lambda$ for the highest value of MSSIM depends on the numerical aperture and the radius of the individual core.

The red circles and cyan diamonds in Fig. 8 represent the value of MSSIM for image transport through disordered ($\Delta \neq 0$) image fibers. The radius of each individual core is chosen from a uniform distribution in the interval $[1.45\ \mu m - \Delta, 1.45\ \mu m + \Delta]$. Therefore, the mean radius is the same as that of the periodic case. $\Delta = 0.3\ \mu m$ for red circles and $\Delta = 0.9\ \mu m$ for cyan diamonds. For each image fiber, the value of MSSIM is calculated using the transported image. The numerical experiment is repeated 30 times for each representation of the disorder (at each $\Lambda$ and $\Delta$), in order to obtain sufficient statistics, where the error-bars in MSSIM signify one standard deviation around the mean MSSIM.

It is observed that at each value of $\Lambda$, the presence of the disorder improves the quality of the transported image. A detailed analysis shows that the maximum value of MSSIM is obtained for $\Lambda = 3.9\ \mu m$ and $\Delta = 0.3\ \mu m$. Although the disordered multicore image fiber explored in here differs from the structure of the p-ALOF, this numerical experiment provides further evidence in support of the claim that the presence of disorder can enhance the quality of the image transport.

Further analysis of the impact of the disorder on the image transport quality and the evolution of MSSIM with the propagation distance in these fibers is provided in the Supplementary Information in Supplementary Fig. S. 1, Supplementary Fig. S. 2, and the related discussions.

**Impact of the wavelength on image quality.** For im-age transport through multicore imaging optical fibers, a longer optical wavelength increases the inter-core coupling strength and therefore lowers the quality of image transport. A similar effect can be observed in Anderson localized optical fibers as the mean localized beam radius has been shown to be larger for a longer wavelength [9].

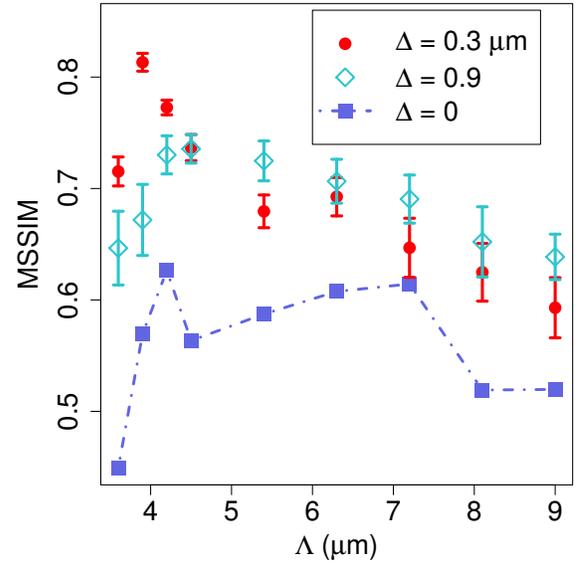

Fig. 8. **Quantitative assessment of image transport quality through disordered fibers.** The MSSIM metric for image quality assessment is compared for periodic image fibers as a function the core periodicity $\Lambda$. The blue squares represent disorder-free ($\Delta = 0$) periodic cores of radius 1.45 $\mu m$. The red circles and cyan diamonds represent disordered ($\Delta \neq 0$) periodic cores of mean radius 1.45 $\mu m$ with uniform random radius variations in the interval $[-0.3\ \mu m, 0.3\ \mu m]$ and $[-0.9\ \mu m, 0.9\ \mu m]$, respectively.

In Fig. 9, a longer wavelength of 633 nm is used for the numerical simulation of the image transport. The results must be compared with Fig. 6. Fig. 9 shows the numerical simulation of transported images related to group 5 of the 1951-AFTT test chart in (a) FIGH-10-350S image fiber, (b) FIGH-10-350S image fiber, and (c) p-ALOF, using the 633 nm wavelength. The quality of the transported image at 633 nm is substantially lower than at 405 nm wavelength for all cases. For p-ALOF, the localization radius was calculated to be smaller than 10 $\mu m$ at 405 nm wavelength and nearly 30 $\mu m$ at 633 nm wavelength [9]. Because the localization radius determines the image transport resolution in a disordered optical fiber, the resolution is nearly three times lower at 633 nm than at 405 nm wavelength. For image fibers, longer wavelength results in a larger modal overlap between the cores, hence increasing core-to-core coupling and blurring of the image. It must be noted that the quality for imaging through the p-AOLF at the wavelength of 633 nm is still higher than that of the multicore im-age fibers at the same wavelength. The MSSIM values are 0.3385 for Fig. 9(a), 0.5396 for Fig. 9(b), 0.6119 for Fig. 9(c), and 0.6509 for Fig. 9(d).

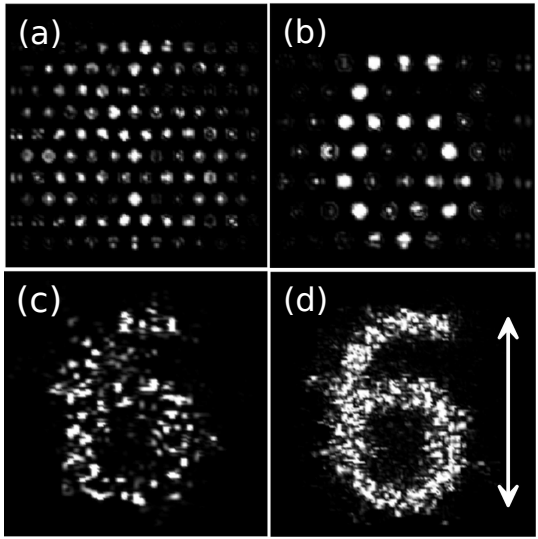

Fig. 9. **Simulation of image transport at a longer wavelength of 633 nm.** Numerical simulation of transported images related to group 5 of the 1951-AFTT test chart in (a) FIGH-10-350S image fiber, (b) FIGH-10-500N image fiber, and (c) p-ALOF. Each fiber is approximately 5 cm long. The simulation wavelength is 633 nm and lower quality imaging is obtained, as expected, when compared with Fig. 6 at 405 nm wavelength. (d) is similar to (c), except an air-glass material with the index difference of 0.5 is assumed instead of 0.1 related to the polymer p-ALOF of (c). The scale bar in (d) is 20 $\mu$m long and the same scale bar can be used for (a), (b), and (c). The MSSIM image quality value for the images are: (a) 0.3385, (b) 0.5396, (c) 0.6119, and (d) 0.6509.

**Impact of the disorder refractive index difference on image quality.** A higher value of the refractive index difference between the random sites in a disordered fiber such as p-ALOF is expected to improve the image transport quality by decreasing the localized beam radius [9]. An ideal disordered image fiber is made from a glass matrix with wavelength-size randomly distributed air-holes to provide a refractive index contrast of 0.5 between high- and low-index sites. The ideal air-hole fill-fraction is at 50% for maximum transverse scattering and minimum localized beam radius [9]. In Fig. 9(d), we show the image transport through a disordered fiber with the same geometrical parameter of p-ALOF, but the refractive index values of the high- and low-index sites are set at 1.5 and 1.0, respectively, for the wavelength of the 633 nm. The quality of image transport in Fig. 9(d) is substantially better than Fig. 9(c).

We note that not only is the average localized beam radius lower in a properly designed glass-air disordered optical fiber compared with a p-ALOF, but there also are fewer sample-to-sample variations in the localized beam radius. Therefore, a higher image transport quality is expected with a more uniform performance across the fiber.

**Signal attenuation along the fiber.** Much lower intrinsic attenuation is expected in silica-based fibers compared with polymer fibers for longer image transport. In fact, 16 cm has been the longest p-ALOF sample that has been successfully used for imaging; transported images of the numbers "1" and "6" from group 3 of the 1951-AFTT test chart are shown in Fig. 10. The maximum length for image transport in the p-ALOF is both a result of optical attenuation and the variations in the thickness of the optical fiber in the draw process. White light has been used for the image transport in Fig. 10, and the output images appear red because the attenuation of the polymer materials used in p-ALOF is minimum for the red color in the visible spectrum. The large variations in the thickness of the optical fiber in the draw process and the accumulated dust and humidity during the week-long stacking of the fibers due to the considerable static electrical charges that build up on the polymer fibers result in a loss of around 0.5 dB/cm at 633 nm. However, it is expected that by assembling the fibers in a clean-room environment and using a more stable draw process, the length of p-ALOF image fibers can be extended to at least several meters.

A glass-air disordered optical fiber would be an ideal solution to the attenuation problem. Moreover, it is possible to obtain high quality fiber-end surfaces by using standard cleaving and polishing equipment in the laboratory. For the glass-air disordered fibers reported in Ref. [21] and shown in Fig. 1(b), transverse Anderson localization can be observed only near the boundary, because their air fill-fraction is 2-5% in the central regions, which is far below the ideal 50%; even near the boundary where transverse Anderson localization is observed, the localized beam radius is too large to be suitable for image transport.

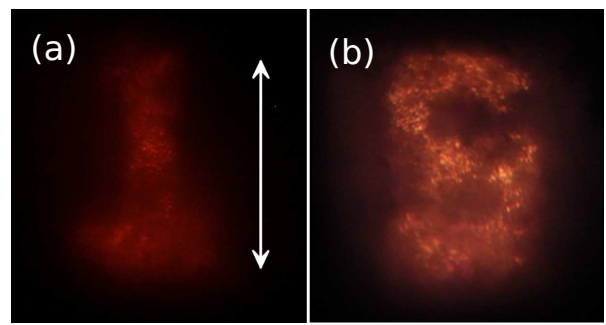

Fig. 10. **Image transport through the disordered fiber over a longer distance.** Transported images of the numbers "1" and "6" from group 3 of the 1951-AFTT test chart through a 16 cm-long p-ALOF sample are shown using a white light source (experimental measurements). The scale bar in (a) is 120 $\mu$m long and the same scale bar can be used for (b).

**Comparison with other advanced fiber-based imaging methods.** Finally, it would be important to compare the advantages and disadvantages of imaging using the disordered optical fibers with other available fiber-based techniques. For example, Choi, et al., [22], recently developed a highly interesting technique to use a single-core multimode optical fiber for endoscopic imaging. They relied on pre- and post-processing techniques to evaluate the transmission matrix of a multimode optical fiber to address the modal dispersion. Also, speckle imaging and turbid lens imaging methods are used to eliminate distortions. The post-processing limits the image acquisition time to one frame per second. In comparison, the disordered fiber provides a one-to-one map between the object and image plane and no additional optics or pre- and post-processing is required to obtain the image. Therefore, image acquisition time is negligible.

Another advantage of the disordered fiber-imaging over the technique presented by Choi, et al., [22], is that their method does not support a fully flexible endoscopic operation due to the variation of the transmission matrix induced by the bending and twisting of the fiber. The beam localization in disordered fiber is very robust, even in very tight bends as shown in Ref. [8], so it can easily operate in a fully flexible endoscopic system.

The main advantage of the method of Ref. [22] over the disordered fiber-imaging is the imaging resolution. They report 12300 effective image pixels over the 200 $\mu m$ diameter core of the fiber, hence a pixel size of $6.4\lambda^2$. Based on our simulations, even the best glass-air disordered fibers can only achieve pixel resolution of larger than $\sim 30\lambda^2$, which is lower than that reported by Choi, et al. On the other hand, the disordered fiber is easily scalable, so it is conceivable the fabricate disordered fibers with millimeter-size cross sections for ultra-wide-area and nearly instantaneous imaging for streaming videos, while the post-processing would make other techniques prohibitively slow. It must also be noted that the disordered fiber is inherently a multimode fiber and the technique employed by Choi, et al., can be applied to the disordered fiber as well. In that case, the localized nature of the beam transport and also the direct knowledge of transported images, albeit at a slightly lower resolution, can potentially reduce the post-processing time in their technique.

In conclusion, we present what is, to the best of our knowledge, the first demonstration of optical image transport using transverse Anderson localization of light in a polymer disordered optical fiber (p-ALOF first reported in Ref. [7]). The image transport quality is comparable with or better than some of the best commercially available multicore image fibers with less pixelation and higher contrast. In practice, the imaging resolution in p-ALOF is limited by the quality of the cleave and polishing of the p-ALOF surface, and the maximum transport distance is limited by the optical attenuation as well as the variations in the thickness of p-ALOF in the draw process. The ultimate disordered image fiber will be made from a glass matrix with wavelength-size randomly distributed air-holes with an air-hole fill-fraction of 50%. The low optical attenuation in glass-air material is essential for transporting images over longer distances than reported here. The large index difference between the glass matrix and the air-holes and 50% air-hole fill-fraction provide maximum scattering in the transverse plane to reduce the localization beam radius and to minimize the width of the imaging point spread function. The large transverse scattering is also responsible for reducing the beam-to-beam variation of the localization radius. A small value of the standard deviation is essential for device applications, as one expects to observe a nearly uniform width of the point spread function across the tip of the image fiber as well as among fiber samples. The reduction in the value of the standard deviation of the localized beam radius is due to the self-averaging behavior observed in the presence of strong scattering [7, 18, 19]. In general, Anderson localization is a statistical problem and the localization happens only strictly when averaged over many elements of a statistically identical ensemble. However, the large transverse scattering results in a strong self-averaging behavior where the localized beam radius of each element of the ensemble is nearly equal to the average localized beam radius; therefore, the standard deviation is small. Further details on the self-averaging property of highly scattering disordered fibers can be found in Ref. [9]. The image transport quality is also independent of the polarization as shown in Supplementary Fig. S. 3 and the related discussion.

Future efforts are focused on designing an air-glass disordered fiber with 50% air-hole fill-fraction, where the average size of the individual air-holes will be using numerical simulations optimized to obtain the minimum beam localization radius. We note that image transport has been investigated previously in random phase-separated glasses [23, 24]. Unlike the p-ALOF, their reported structure lacks longitudinal invariance. Fabrication of an ideal air-glass disordered fiber is very challenging using the stack-and-draw techniques, considering that 100,000 or more elements must be used and the airholes must remain open after the draw. It may be possible to achieve this using a lower number of sites and by restacking/redrawing to obtain the required submicron resolution of the disorder. It may also be possible to use porous glass such as that reported in Ref. [21], albeit with a higher air-hole fill-fraction of nearly 50%. Future efforts will also include revisiting image transport in random phase-separated glasses and its relationship to transverse Anderson localization.

## Methods

**Fabrication of the fiber.** 40,000 pieces of PMMA and 40,000 pieces of PS fibers were randomly mixed; each fiber was 8 inches long with an approximate diameter of ∼200 $\mu m$. The mixture was fused together and redrawn to a fiber with a nearly square profile with an approximate side width of 250-$\mu m$. While some of the original randomly mixed fibers might have crossed over each other during the assembly and redraw process, the large redraw ratio guarantees that the refractive index profile remains relatively unchanged along the fibers in our experiments.

The image fibers used for comparison with p-ALOF are from Fujikura (FIGH-10-350S and FIGH-10-500N) and are distributed by Myriad Fiber Imaging Technology, Incorporated.

**Experiments.** For imaging, we chose fiber samples of nearly 5 cm long and imaged a section of 1951 U.S. Air Force resolution test target (R1DS1N from Thorlabs) by directly butt-coupling the test target to the polished input end of the p-ALOF. The test target was directly illuminated by a diode laser at the wavelength of 405 nm, by a He-Ne at the wavelength of 633 nm, or by white light. The near-field output was imaged using 40x and 60x objectives onto a lensed CCD camera.

**Numerical simulations.** The images were propagated along the fiber by numerically solving the wave propagation equation equation described in Ref. [9]. The illuminated image "UWM" and number "6" were created using the GIMP image editor and used as the input. The disordered refractive index profile was created by randomly assigning the refractive indexes of $n_1$ and $n_2$ to the different sites. The sites are square with the side width of 0.9-$\mu$m.

**Assessment of the image quality.** Mean structural similarity index (MSSIM) is used in this manuscript as a quantitative measure to closely approximate the perceived image quality by human eye (see the Supplementary Information for more details) [25,26]. MSSIM is defined based on the structural similarity index (SSIM). In SSIM, the quality of the transported images are compared with the distortion free initial images that are launched into the fibers, by correlating the local pixel intensity patterns in different regions of the two images. The source of distortion in the transported images in the multicore image fibers studied here are the pixelation and coupling of light to adjacent cores.

SSIM is used to compare the local image patches x and y from the same locations of the two images [26]. A simplified form of SSIM index is calculated using:

$$\text{SSIM}(x,y) = \frac{(2\mu_x \mu_y + C_1)(2\sigma_{xy} + C_2)}{(\mu_x^2 + \mu_y^2 + C_1)(\sigma_x^2 + \sigma_y^2 + C_2)}, \quad (1)$$

where $\mu_x$ and $\mu_y$ are local sample means of x and y, $\sigma_x$ and $\sigma_y$ are local sample standard deviation of x and y, and $\sigma_{xy}$ is the cross correlation of samples x and y after subtracting their mean. $C_1$ and $C_2$ are positive numbers to stabilize the SSIM in cases of near zero mean, standard deviation or cross correlation. $C_1$ and $C_2$ are defined as $(K_1 L)^2$ and $(K_2 L)^2$ where L is the dynamic range of the pixel intensities and $K_1 \ll 1$, $K_2 \ll 1$. MSSIM index is calculated by averaging the SSIM over the image patches

$$\text{MSSIM} = \frac{1}{N} \sum_{j=1}^{N} \text{SSIM}(x_j, y_j), \quad (2)$$

where N is the number of local patches of the image and $x_j$ and $y_j$ are the image contents at the j*th* patch.

### Acknowledgments

This research is supported by grant number 1029547 from the National Science Foundation. The authors would like to acknowledge David J. Welker from Paradigm Optics Inc. for providing the initial polymer fiber segments and the re-drawing of the final p-ALOF.

### Author Contributions

S.K. and A.M. wrote the manuscript and all authors contributed to its final editing. S.K. contributed to the fabrication of the p-ALOF, S.K. and R.F. conducted all experiments; S.K. analyzed the experimental data and prepared the figures. S.K. carried out all the numerical simulations. T.H. and J.B. fabricated the glass-air disordered fiber. A.M. and K.K. conceived the original idea of making an ALOF. A.M. led the project and supervised all aspects of the work.